\documentclass[conference]{IEEEtran}
\usepackage{cite}
\usepackage{amsmath,amssymb,amsfonts}
\usepackage{graphicx}
\usepackage{textcomp}
\usepackage{xcolor}
\usepackage{array}
\usepackage{tikz}
\usetikzlibrary{arrows.meta,positioning}
\def\BibTeX{{\rm B\kern-.05em{\sc i\kern-.025em b}\kern-.08em
    T\kern-.1667em\lower.7ex\hbox{E}\kern-.125emX}}

\begin{document}
\bstctlcite{IEEEexample:BSTcontrol}

\title{Optimal Energy and LLM Task Scheduling for Multiple Cloud Data Centers}

\author{\IEEEauthorblockN{Zhongqi Zhao\IEEEauthorrefmark{1}\IEEEauthorrefmark{2},
Lei Fan\IEEEauthorrefmark{1}\IEEEauthorrefmark{2}, and
Zhu Han\IEEEauthorrefmark{1}}
\IEEEauthorblockA{\IEEEauthorrefmark{1}\textit{Department of Electrical and Computer Engineering, University of Houston, Houston, TX, USA}}
\IEEEauthorblockA{\IEEEauthorrefmark{2}\textit{Department of Engineering Technology, University of Houston, Houston, TX, USA}}}
\maketitle

\begin{abstract}
This paper studies coordinated Large Language Model (LLM) training-task scheduling and facility operation for geographically distributed cloud data centers (CDCs). The central question is how two scheduling flexibilities---cross-CDC task transfer and task reordering---affect service and economic outcomes when remote execution remains subject to compute, bandwidth, energy, and zone-temperature constraints. We formulate a mixed-integer optimization model that jointly determines task outcomes, heterogeneous-node assignment, execution timing, pre-execution data transfer, node activation, and facility operation. A deterministic 24-hour case study contains 3 CDCs, 15 heterogeneous graphics processing unit (GPU) nodes, 4 thermal zones per CDC, and 103 tasks. In this instance, enabling transfer increases net profit by 57.9\% relative to the otherwise matched no-transfer variant and increases completed tasks from 76 to 103. The more restrictive no-transfer FIFO-prefix variant completes 14 tasks and yields negative profit. The proposed schedule transfers 39 tasks, achieves 70.3\% node utilization, and maintains zone temperatures within 24$^\circ$C--27$^\circ$C. These fixed-instance results indicate that cross-CDC transfer and task-reordering flexibility can improve the modeled profit and task-completion outcomes under the studied conditions.
\end{abstract}


\section{Introduction} \label{par:Intro}
Training Large Language Models (LLMs) requires substantial computation, data movement, and storage input/output (I/O), and these workloads are commonly executed in cloud data centers (CDCs) \cite{qian2024alibaba,gunnam2025llm}. In a geographically distributed deployment, an operator must decide not only when and where each training task runs, but also whether its data should be transferred to another CDC. That placement decision changes node use, communication demand, heat generation, and facility energy requirements.

Prior work has addressed several parts of this problem. Cooling-control and digital-twin methods target facility energy \cite{zhang2022smart}, while learning-based approaches schedule or migrate data-center tasks \cite{lou2022energy}. Recent studies also consider multi-cloud task scheduling, data-center network management, and geo-distributed job-and-data placement \cite{tang2025llm,srikandabala2025optimization,ding2026energy}. These advances establish that geographic placement, communication, and facility operation matter. They do not, however, determine how much value arises from cross-CDC transfer and how much arises from relaxing a restrictive task order when both mechanisms are evaluated under the same physical and economic constraints.

LLM training tasks are both computation- and data-intensive. Their placement therefore cannot be based on processor availability alone: a remote task consumes outbound and inbound bandwidth, and its data must arrive before execution begins. CDCs also differ in node composition, available capacity, and operating conditions. A remote assignment can improve the modeled objective only when its placement benefit offsets transfer cost and remains feasible under these constraints \cite{jaiswal2025serving,ding2026energy}.

Task placement is also coupled to facility operation. Activating a node changes computing energy and zone heat, which affects air-conditioning demand and the site energy balance. Storage, renewable supply, and grid purchases further constrain feasible operation. Carbon-aware scheduling across AI data-center parks similarly shows that workload flexibility must be interpreted together with site conditions \cite{han2026carbon}. Here, these facility constraints provide the feasibility context for comparing the two scheduling flexibilities; they are not treated as an independent claim of universal energy efficiency.

Motivated by this distinction, we ask a question regarding to a fixed time horizon: under one coupled compute--communication--facility model, what changes when cross-CDC transfer is disabled, and what changes when task reordering is additionally replaced by a FIFO-prefix restriction? The comparison is diagnostic rather than statistical. It separates the two modeled flexibility mechanisms while holding the task set, node set, CDC assignments, rewards, penalties, and energy parameters fixed.

The main contributions of this paper are summarized as follows:
\begin{itemize}
    \item We formulate a multi-CDC optimization model that couples task outcomes and execution timing with heterogeneous-node assignment and bandwidth-constrained, pre-execution task transfer.
    \item We place these decisions within modeled node-energy, zone-temperature, air-conditioning, storage, renewable-supply, and grid-purchase constraints so that each scheduling variant is evaluated under the same cross-layer feasibility conditions.
    \item We use a fixed-instance comparison of the proposed, No Transfer (No TF), and No Transfer with FIFO-prefix order (No TF+FIFO) variants to separate the effects of cross-CDC transfer and task-reordering flexibility on profit, task completion, concurrency, and node utilization.
\end{itemize}

The rest of this paper is organized as follows. Section \ref{par:DF} presents the problem definition and mathematical formulation. Section \ref{par:EXP} reports the numerical results and discusses the scheduling and energy-management behavior of the optimized solution. Section \ref{par:C} concludes the paper.

\section{Problem Definition and Formulations} 
\label{par:DF}
This section presents a mixed-integer programming (MIP) formulation for net operating profit under LLM training-task scheduling, data transfer, and facility constraints.
We consider a set $\mathcal{I}$ of CDCs located in different regions. Each task $j\in\mathcal{J}$ has a fixed home CDC denoted by $h_j\in\mathcal{I}$. Tasks are classified exogenously as non-time-sensitive (NTS), time-sensitive (TS), or already running before the scheduling horizon. The TS class represents users who selected priority service before the optimization begins; priority purchase is not a decision variable. Table~\ref{table1:LLM_task_pool} illustrates the initial task pools.

\begin{table}[tbp]
	\caption{LLM task pools and electricity-price coefficients}
	\label{table1:LLM_task_pool}
	\centering
	\begin{tabular}{c|c|c}
		\textbf{CDC} & \textbf{LLM Task Pool} & \textbf{Price (NMU/kWh)} \\
		\hline
		$\text{CDC}_1$ & $\{1:\text{LLM}^{\text{w}}_{1,1} \quad \cdots \quad m:\text{LLM}^{\text{NTS}}_{1,m}\}$ & $\boldsymbol{\beta}^{\mathrm{e,g}}_{1,\mathcal{T}}$ \\
		$\text{CDC}_2$ & $\{1:\text{LLM}^{\text{NTS}}_{2,1} \quad \cdots \quad p:\text{LLM}^{\text{NTS}}_{2,p}\}$ & $\boldsymbol{\beta}^{\mathrm{e,g}}_{2,\mathcal{T}}$ \\
		$\vdots$ & $\vdots \quad \cdots \quad \vdots$ & $\vdots$ \\
		$\text{CDC}_N$ & $\{1:\text{LLM}^{\text{TS}}_{N,1} \quad \cdots \quad q:\text{LLM}^{\text{TS}}_{N,q}\}$ & $\boldsymbol{\beta}^{\mathrm{e,g}}_{N,\mathcal{T}}$ \\ \hline
	\end{tabular}
\end{table}

\begin{figure}[tbp]
    \centering
    \begin{tikzpicture}[
        x=1cm,y=1cm,
        >=Latex,
        cdc/.style={draw,rounded corners=2pt,thick,minimum width=3.25cm,minimum height=2.45cm},
        zone/.style={draw,minimum width=1.47cm,minimum height=0.72cm,align=center,font=\scriptsize},
        facility/.style={draw,rounded corners=1.5pt,fill=gray!10,minimum width=3.25cm,minimum height=0.43cm,align=center,font=\scriptsize},
        flow/.style={->,thick}
    ]
        \node[cdc] (cdc1) at (0,0) {};
        \node[font=\small\bfseries,anchor=north] at (cdc1.north) {CDC $i$};
        \node[zone] at (-0.78,0.38) {$z_1$\\$\mathcal{N}_{i,1}$};
        \node[zone] at ( 0.78,0.38) {$z_2$\\$\mathcal{N}_{i,2}$};
        \node[zone] at (-0.78,-0.45) {$z_3$\\$\mathcal{N}_{i,3}$};
        \node[zone] at ( 0.78,-0.45) {$z_4$\\$\mathcal{N}_{i,4}$};
        \node[font=\tiny,anchor=south] at (0,-1.12) {heterogeneous GPU nodes and zone HVAC};
        \node[facility,below=0.10cm of cdc1] (fac1)
        {grid + renewable + battery};

        \node[cdc] (cdc2) at (4.80,0) {};
        \node[font=\small\bfseries,anchor=north] at (cdc2.north) {CDC $k$};
        \node[zone] at (4.02,0.38) {$z_1$\\$\mathcal{N}_{k,1}$};
        \node[zone] at (5.58,0.38) {$z_2$\\$\mathcal{N}_{k,2}$};
        \node[zone] at (4.02,-0.45) {$z_3$\\$\mathcal{N}_{k,3}$};
        \node[zone] at (5.58,-0.45) {$z_4$\\$\mathcal{N}_{k,4}$};
        \node[font=\tiny,anchor=south] at (4.80,-1.12) {heterogeneous GPU nodes and zone HVAC};
        \node[facility,below=0.10cm of cdc2] (fac2)
        {grid + renewable + battery};

        \draw[flow]
        ([yshift=6mm]cdc1.east) to[out=38,in=142]
        node[above,sloped,fill=white,inner sep=1pt,font=\tiny] {task-data transfer}
        ([yshift=6mm]cdc2.west);
        \draw[flow]
        ([yshift=-6mm]cdc2.west) to[out=-142,in=-38]
        node[below,sloped,fill=white,inner sep=1pt,font=\tiny] {remote assignment}
        ([yshift=-6mm]cdc1.east);
        \node[font=\scriptsize,above=0.10cm of cdc1] {home tasks: $h_j=i$};
        \node[font=\scriptsize,above=0.10cm of cdc2] {home tasks: $h_j=k$};
    \end{tikzpicture}
    \caption{Multi-CDC problem layout. Each CDC contains heterogeneous nodes partitioned into thermal zones; remote execution requires inter-CDC task-data transfer, while computing and HVAC decisions share the local facility-energy balance.}
    \label{fig:cdc_layout}
\end{figure}
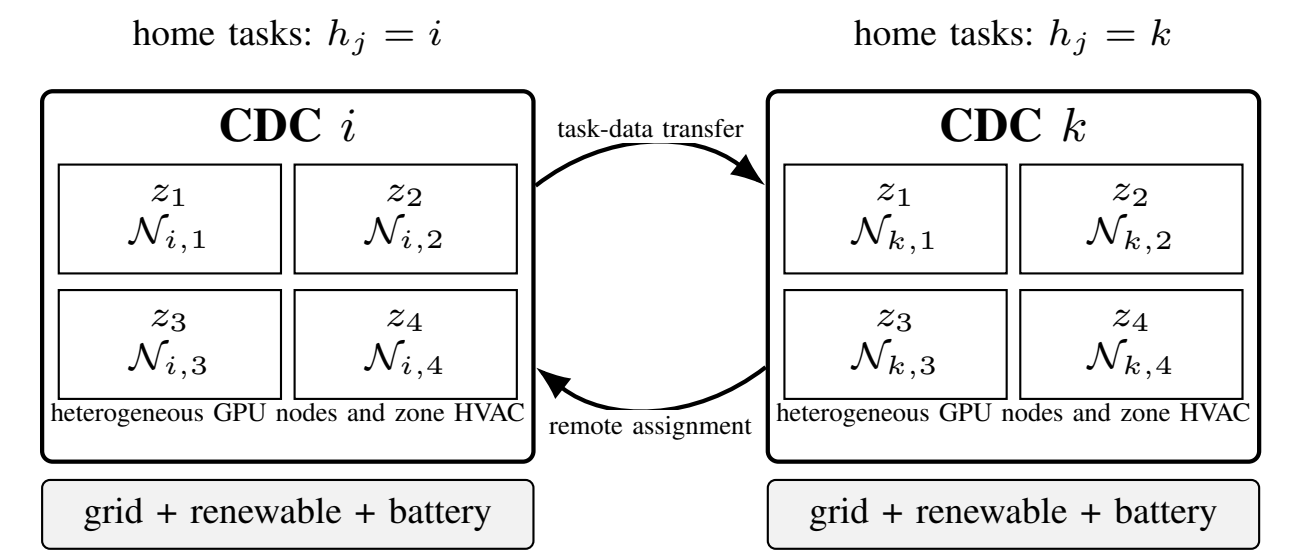

Each CDC contains heterogeneous computing nodes distributed across thermal zones. Let $\mathcal{N}_{i,z}$ be the nodes in zone $z$ of CDC $i$, and let $\ell_n\in\mathcal{I}$ denote the CDC containing node $n$. Fig.~\ref{fig:cdc_layout} illustrates the resulting multi-CDC hierarchy and the coupling between task transfer, zone-level computing, and facility operation.

Given task classes, home CDCs, processing times, deadlines, data volumes, economic coefficients, node limits, and facility profiles, the model determines each task outcome, execution node and interval, required transfer, and CDC energy dispatch. A remote assignment is feasible only when its data reach the destination before startup and both CDCs satisfy their communication limits. The output is therefore a joint task, transfer, node, storage, HVAC, and grid schedule rather than an unconstrained geographic placement.

\subsection{Cloud Data Center Modeling Overview}
We formulate the multi-CDC coordination problem over $t \in \mathcal{T}=\{1,2,\ldots,T\}$ with step length $\Delta t$. The optimizer jointly determines task status, execution node, running trajectory, inter-CDC data transfer, battery charging/discharging, HVAC operation, and grid purchase. Indices $i$, $j$, $n$, $z$, and $t$ denote CDC, task, node, zone, and time slot; sets $\mathcal{I}$, $\mathcal{J}$, $\mathcal{N}$, $\mathcal{Z}$, and $\mathcal{T}$ collect them. Binary variables are mainly $\{x,u,v,z,y\}$; other lowercase symbols denote continuous decision variables. Uppercase symbols and Greek letters denote parameters/constants, except temperature variables such as $T^{\text{Zone}}_{i,z,t}$ and $T^{\text{AC}}_{i,z,t}$.

The complete problem is summarized as follows:
\begin{align}
    \max\ & c^{\mathrm{profit}} - c^{\mathrm{loss}} - c^{\mathrm{Hold}} - c^{\mathrm{TF}} - \sum_{t \in \mathcal{T}} c^{\mathrm{Op}}_{t}, \label{eq1:obj}\\
    \text{s.t.}\ & c^{\mathrm{profit}} = \sum_{j \in \mathcal{J}} C^{\mathrm{profit}}_{j}x^{\mathrm{Done}}_{j}, \label{eq2:profit}\\
    & c^{\mathrm{loss}} = \sum_{j \in \mathcal{J}} C^{\mathrm{loss}}_{j}x^{\mathrm{Abort}}_{j}, \label{eq3:loss}\\
    & c^{\mathrm{Hold}} = \sum_{j \in \mathcal{J}} C^{\mathrm{Hold}}_{j}x^{\mathrm{Hold}}_{j}, \label{eq3:hold}\\
    & c^{\mathrm{TF}} = \sum_{j \in \mathcal{J}} C^{\mathrm{transfer}}_{j}x^{\mathrm{TF}}_{j}, \label{eq4:TF}\\
    & c^{\mathrm{Op}}_{t} = \sum_{i \in \mathcal{I}}
    \beta^{\mathrm{e,g}}_{i,t} e^{\mathrm{Grid}}_{i,t},
    \ \forall t \in \mathcal{T}. \label{eq5:op}
\end{align}
The objective in \eqref{eq1:obj} maximizes net profit. Variables $c^{\mathrm{profit}}$, $c^{\mathrm{loss}}$, $c^{\mathrm{Hold}}$, $c^{\mathrm{TF}}$, and $c^{\mathrm{Op}}_{t}$ are completed-task revenue, abortion penalty, holding penalty, transfer cost, and grid-purchase cost, all in normalized monetary units (NMU). Parameters $C^{\mathrm{profit}}_{j}$, $C^{\mathrm{loss}}_{j}$, $C^{\mathrm{Hold}}_{j}$, and $C^{\mathrm{transfer}}_{j}$ are in NMU/task. Variable $e^{\mathrm{Grid}}_{i,t}$ is grid-purchased energy in kWh/slot, $\beta^{\mathrm{e,g}}_{i,t}$ is the electricity-price coefficient in NMU/kWh, and $c^{\mathrm{Op}}_{t}$ is consequently in NMU/slot.

\subsection{LLM Training Task Scheduling}
The scheduling block includes four parts: final task states, execution assignment and temporal consistency, node-level capacity coupling, and cross-CDC transfer feasibility.

\subsubsection{Task States and Assignment}
\begin{align}
    & x^{\mathrm{Done}}_{j}+x^{\mathrm{Abort}}_{j}+x^{\mathrm{Hold}}_{j}=1, \ \forall j \in \mathcal{J}, \label{eq6:job_State_set}\\
    & x^{\mathrm{TF}}_{j}=0, \ \forall j \in \mathcal{J}^{\mathrm{w}}, \label{eq7:notf}\\
    & x^{\mathrm{Hold}}_{j}=0, \ \forall j \in \mathcal{J}^{\mathrm{w}} \cup \mathcal{J}^{\mathrm{TS}}, \label{eq8:nohold}\\
    & x^{\mathrm{Abort}}_{j}=0, \ \forall j \in \mathcal{J}^{\mathrm{NTS}}\cup\mathcal{J}^{\mathrm{w}}, \label{eq9:noabort}\\
    & \sum_{n \in \mathcal{N}} x^{\mathrm{task}}_{j,n}=x^{\mathrm{Done}}_{j}, \ \forall j \in \mathcal{J}. \label{eq10:assign}
\end{align}
Equation \eqref{eq6:job_State_set} assigns each task to exactly one base state. Binary variables $x^{\mathrm{Done}}_{j}$, $x^{\mathrm{Abort}}_{j}$, and $x^{\mathrm{Hold}}_{j}$ indicate completion, abortion, and holding; $x^{\mathrm{TF}}_{j}$ indicates remote execution; and $x^{\mathrm{task}}_{j,n}$ indicates assignment to node $n$. Sets $\mathcal{J}^{\mathrm{w}}$, $\mathcal{J}^{\mathrm{TS}}$, and $\mathcal{J}^{\mathrm{NTS}}$ are already-running, time-sensitive, and non-time-sensitive tasks. Constraints \eqref{eq7:notf}--\eqref{eq10:assign} enforce state eligibility and assignment consistency.

\subsubsection{Temporal Scheduling Constraints}
{\footnotesize
\begin{align}
    & u^{\mathrm{task}}_{j,n,t} \le x^{\mathrm{task}}_{j,n}, \ \forall j \in \mathcal{J}, \forall n \in \mathcal{N}, \forall t \in \mathcal{T}, \label{eq11:runassign}\\
    & u^{\mathrm{task}}_{j,n,t-1}-u^{\mathrm{task}}_{j,n,t}-v^{\mathrm{SD}}_{j,n,t}+v^{\mathrm{SU}}_{j,n,t}=0, \ \forall j \in \mathcal{J}, \forall n \in \mathcal{N}, \forall t \in \mathcal{T}, \label{eq12:onoff}\\
    & v^{\mathrm{SD}}_{j,n,t}+v^{\mathrm{SU}}_{j,n,t} \le 1, \ \forall j \in \mathcal{J}, \forall n \in \mathcal{N}, \forall t \in \mathcal{T}, \label{eq13:jobsusd}\\
    & \sum_{n \in \mathcal{N}} \sum_{t \in \mathcal{T}} u^{\mathrm{task}}_{j,n,t} = \sum_{n \in \mathcal{N}} T^{\mathrm{task}}_{j,n} x^{\mathrm{task}}_{j,n}, \ \forall j \in \mathcal{J}^{\mathrm{NTS}} \cup \mathcal{J}^{\mathrm{TS}}, \label{eq14:duration}\\
    & \sum_{\tau=t}^{\min\{t+T^{\mathrm{task}}_{j,n}-1,T\}} u^{\mathrm{task}}_{j,n,\tau} \ge T^{\mathrm{task}}_{j,n} v^{\mathrm{SU}}_{j,n,t}, \label{eq15:nonpreemptive}\\
    & \qquad \forall j \in \mathcal{J}^{\mathrm{NTS}} \cup \mathcal{J}^{\mathrm{TS}}, \ \forall n \in \mathcal{N}, \ \forall t=1,\ldots,T-T^{\mathrm{task}}_{j,n}+1, \notag\\
    & \sum_{n \in \mathcal{N}} \sum_{t \in \mathcal{T}} v^{\mathrm{SU}}_{j,n,t}=x^{\mathrm{Done}}_{j}, \ \forall j \in \mathcal{J}^{\mathrm{NTS}} \cup \mathcal{J}^{\mathrm{TS}}, \label{eq16:onestart}\\
    & \sum_{n \in \mathcal{N}} \sum_{t=1}^{T^{\mathrm{TS}}_{j}} u^{\mathrm{task}}_{j,n,t}=\sum_{n \in \mathcal{N}} T^{\mathrm{task}}_{j,n}x^{\mathrm{task}}_{j,n}, \ \forall j \in \mathcal{J}^{\mathrm{TS}}, \label{eq17:tsdeadline}\\
    & u^{\mathrm{task}}_{j,N^{\mathrm{w}}_{j},0}=1, \ \forall j \in \mathcal{J}^{\mathrm{w}}, \label{eq18:workinginit1}\\
    & u^{\mathrm{task}}_{j,n,0}=0, \ \forall j \in \mathcal{J}^{\mathrm{w}}, \label{eq19:workinginit0}\\
    & \qquad \forall n \in \mathcal{N}, \ n \neq N^{\mathrm{w}}_{j}. \notag\\
    & u^{\mathrm{task}}_{j,n,0}=0,\ \forall j\in\mathcal{J}^{\mathrm{NTS}}\cup\mathcal{J}^{\mathrm{TS}},\ \forall n\in\mathcal{N}, \label{eq19:idleinit}\\
    & \sum_{t\in\mathcal{T}}u^{\mathrm{task}}_{j,N^{\mathrm{w}}_j,t}
      =T^{\mathrm{rem}}_j x^{\mathrm{Done}}_j,\ \forall j\in\mathcal{J}^{\mathrm{w}}. \label{eq19:workingduration}
\end{align}
}
Binary variable $u^{\mathrm{task}}_{j,n,t}$ indicates that task $j$ runs on node $n$ at time $t$, while $v^{\mathrm{SU}}_{j,n,t}$ and $v^{\mathrm{SD}}_{j,n,t}$ indicate startup and shutdown. Parameters $T^{\mathrm{task}}_{j,n}$, $T^{\mathrm{TS}}_{j}$, $N^{\mathrm{w}}_{j}$, and $T^{\mathrm{rem}}_j$ are processing duration in integer slots, deadline slot, initial node, and remaining duration of an already-running task. Constraints \eqref{eq11:runassign}--\eqref{eq19:workingduration} enforce assignment consistency, startup/shutdown logic, non-preemptive execution, deadlines, and initial/remaining-work conditions.

\subsubsection{Node Capacity and Node Power Coupling}
\begin{align}
    & \sum_{j \in \mathcal{J}} r_{j,n} u^{\mathrm{task}}_{j,n,t} \le \overline{R}_{n}, \ \forall n \in \mathcal{N}, \forall t \in \mathcal{T}, \label{eq20:nodecap}\\
    & \sum_{j \in \mathcal{J}} u^{\mathrm{task}}_{j,n,t} \le 1, \ \forall n \in \mathcal{N}, \forall t \in \mathcal{T}, \label{eq21:nooverlap}\\
    & e^{\mathrm{CDC}}_{i,t}=\sum_{n \in \mathcal{N}_{i}} e^{\mathrm{Node}}_{n,t}, \ \forall i \in \mathcal{I}, \forall t \in \mathcal{T}, \label{eq22:edc}\\
    & e^{\mathrm{Node}}_{n,t}=\left(P^{\mathrm{idle}}_{n} z_{n,t}+\sum_{j \in \mathcal{J}} P^{\mathrm{task}}_{j,n}u^{\mathrm{task}}_{j,n,t}\right)\Delta t, \label{eq23:nodepower}\\
    & \qquad \forall n \in \mathcal{N}, \ \forall t \in \mathcal{T}, \notag\\
    & u^{\mathrm{task}}_{j,n,t} \le z_{n,t}, \ \forall j \in \mathcal{J}, \forall n \in \mathcal{N}, \forall t \in \mathcal{T}. \label{eq24:jobnodeon}
\end{align}
Here, $r_{j,n}$ is the resource demand, $e^{\mathrm{CDC}}_{i,t}$ and $e^{\mathrm{Node}}_{n,t}$ are CDC- and node-level computing energy, and $z_{n,t}$ is the node ON/OFF state. Parameters $\overline{R}_{n}$, $P^{\mathrm{idle}}_{n}$, and $P^{\mathrm{task}}_{j,n}$ are node capacity, idle power, and task-induced power; $\mathcal{N}_{i}$ is the node set in CDC $i$. Constraints \eqref{eq20:nodecap}--\eqref{eq24:jobnodeon} enforce capacity, no-overlap, energy aggregation, node-energy accounting, and ON/OFF coupling.

\subsubsection{Transferred Tasks and Communication Constraints}
\begin{align}
    & x^{\mathrm{TF}}_{j}=x^{\mathrm{Done}}_{j}-\sum_{\{n \in \mathcal{N}\mid \mathcal{I}^{N}(n)=\mathcal{I}^{J}(j)\}} x^{\mathrm{task}}_{j,n}, \ \forall j \in \mathcal{J}, \label{eq25:tfdef}\\
    & D_{j}x^{\mathrm{TF}}_{j}=\sum_{t \in \mathcal{T}} y^{\mathrm{bw,out}}_{j,t}, \ \forall j \in \mathcal{J}^{\mathrm{NTS}}, \label{eq26:tfnts}\\
    & D_{j}x^{\mathrm{TF}}_{j}=\sum_{t=1}^{T^{\mathrm{TS}}_{j}} y^{\mathrm{bw,out}}_{j,t}, \ \forall j \in \mathcal{J}^{\mathrm{TS}}, \label{eq27:tfts}\\
    & y^{\mathrm{bw,out}}_{j,t} \le \overline{Y}^{\mathrm{bw,out}}_{j}x^{\mathrm{TF}}_{j}, \ \forall j \in \mathcal{J}, \forall t \in \mathcal{T}, \label{eq28:bwout}\\
    & \sum_{\{j \in \mathcal{J}\mid \mathcal{I}^{J}(j)=i\}} y^{\mathrm{bw,out}}_{j,t} \le \overline{\mathrm{BW}}^{\mathrm{out}}_{i}, \ \forall i \in \mathcal{I}, \forall t \in \mathcal{T}, \label{eq29:bwoutdc}\\
    & y^{\mathrm{bw,out}}_{j,t}=\sum_{\{n \in \mathcal{N}\mid \mathcal{I}^{N}(n)\neq \mathcal{I}^{J}(j)\}} y^{\mathrm{bw,in}}_{j,n,t}, \ \forall j \in \mathcal{J}, \forall t \in \mathcal{T}, \label{eq30:bwbalance}\\
    & y^{\mathrm{bw,in}}_{j,n,t} \le \overline{Y}^{\mathrm{bw,in}}_{j,n}x^{\mathrm{task}}_{j,n}, \ \forall j \in \mathcal{J}, \forall n \in \mathcal{N}, \forall t \in \mathcal{T}, \label{eq31:bwin}\\
    & \sum_{\{j \in \mathcal{J},\,n \in \mathcal{N}_{i}\}} y^{\mathrm{bw,in}}_{j,n,t} \le \overline{\mathrm{BW}}^{\mathrm{in}}_{i}, \ \forall i \in \mathcal{I}, \forall t \in \mathcal{T}, \label{eq32:bwindc}\\
    & \sum_{\tau=1}^{t} y^{\mathrm{bw,in}}_{j,n,\tau} \ge D_{j}v^{\mathrm{SU}}_{j,n,t+1}, \label{eq33:databeforerun}\\
    & \qquad \forall j \in \mathcal{J}, \ \forall t=1,\ldots,T-1, \notag\\
    & \qquad \forall n \in \mathcal{N}:\mathcal{I}^{N}(n)\neq \mathcal{I}^{J}(j). \notag\\
    & v^{\mathrm{SU}}_{j,n,1}=0,\ \forall j\in\mathcal{J},\
      \forall n\in\mathcal{N}:\ell_n\neq h_j. \label{eq33:nostartslotone}
\end{align}
Equation \eqref{eq25:tfdef} defines remote execution. Because the supplied data do not identify a physical byte scale, we use normalized data units (NDU): $D_j$ is in NDU, while $y^{\mathrm{bw,out}}_{j,t}$ and $y^{\mathrm{bw,in}}_{j,n,t}$ are outbound and inbound transfer volumes in NDU/slot. Parameters $\overline{Y}^{\mathrm{bw,out}}_{j}$, $\overline{Y}^{\mathrm{bw,in}}_{j,n}$, $\overline{\mathrm{BW}}^{\mathrm{out}}_{i}$, and $\overline{\mathrm{BW}}^{\mathrm{in}}_{i}$ are likewise in NDU/slot. Constraints \eqref{eq26:tfnts}--\eqref{eq33:nostartslotone} enforce full transmission, bandwidth caps, flow balance, and data arrival before remote startup. The case study assumes no pre-positioned remote task data, so a remote task cannot start in slot 1.

\begin{table}[t]
\caption{Implemented computing-node inventory}
\label{table:node_specs}
\centering
\begin{tabular}{|l|c|}
\hline
\textbf{Node Type} & \textbf{Number of Nodes} \\
\hline
DGX1 & 4 \\ \hline
DGXA100 & 4 \\ \hline
DGXA100s & 2 \\ \hline
DGXA200 & 2 \\ \hline
DGXH100 & 3 \\
\hline
\end{tabular}
\end{table}

\subsection{Facility Energy and Zone-Temperature Constraints}
We adapt the data-center energy model in \cite{zhao2023optimal}. For each one-hour slot, the site energy balance is
\begin{equation}
\begin{aligned}
e^{\mathrm{Grid}}_{i,t}+e^{\mathrm{S}}_{i,t}+p^{\mathrm{dis}}_{i,t}\Delta t
=e^{\mathrm{HVAC}}_{i,t}+e^{\mathrm{CDC}}_{i,t}
+p^{\mathrm{ch}}_{i,t}\Delta t .
\end{aligned}
\label{cost_decomp}
\end{equation}
Here, $p^{\mathrm{ch}}_{i,t}$ and $p^{\mathrm{dis}}_{i,t}$ are battery charging and discharging power. The storage-state equation includes charging and discharging efficiencies, and renewable energy in excess of modeled demand may be curtailed.

The MIP uses an affine discrete-time zone model,
\begin{equation}
\begin{aligned}
T^{\mathrm{Zone}}_{i,z,t}
={}&T^{\mathrm{Zone}}_{i,z,t-1}
+\sum_{z'\in\mathrm{adj}(i,z)}
a_{i,z',z}\!\left(T^{\mathrm{Zone}}_{i,z',t-1}
-T^{\mathrm{Zone}}_{i,z,t-1}\right)\\
&+b_{i,z}\theta_{i,z,t}-q^{\mathrm{AC}}_{i,z,t},\\
\theta_{i,z,t}={}&\sum_{n\in\mathcal{N}_{i,z}}e^{\mathrm{Node}}_{n,t},
\quad \forall i,z,t .
\end{aligned}
\label{Zone_Temperature_Model}
\end{equation}
Coefficients $a_{i,z',z}$ and $b_{i,z}$ absorb the one-hour discretization and zone thermal parameters; $q^{\mathrm{AC}}_{i,z,t}$ is the modeled temperature-reduction term linked to HVAC energy and bounded supply conditions. This form assigns each node's heat contribution only to its own zone and keeps the printed model affine. The case study enforces the reported zone-temperature bounds. The supplied result package does not document a cyclic terminal battery condition, so the grid-purchase results below are interpreted only for this 24-h instance.

\begin{figure*}[t]
    \centering
    \includegraphics[width=\linewidth]{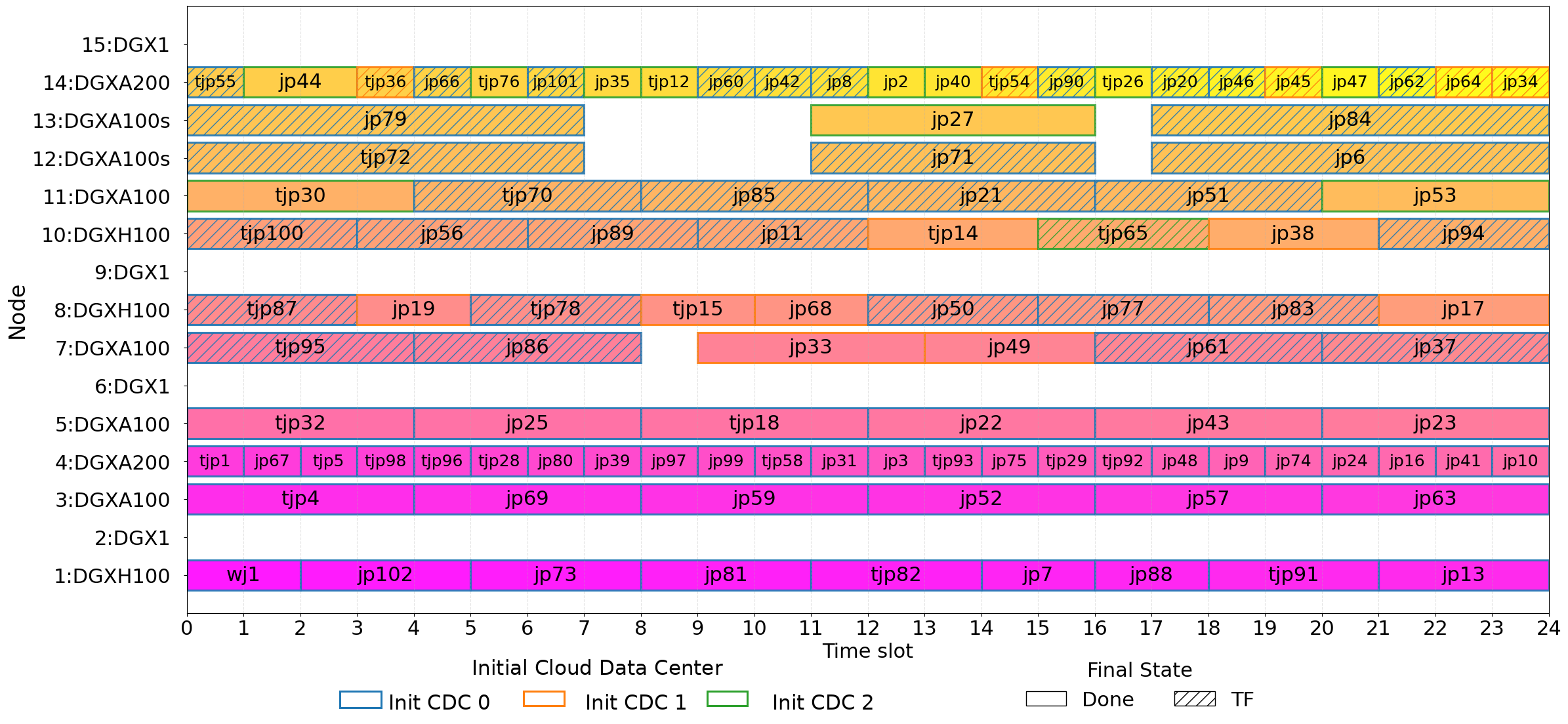}
    \caption{LLM task schedule across heterogeneous nodes.}
    \label{fig1:task_state}
\end{figure*}

\section{Numerical Validation}\label{par:EXP}
We solve the deterministic case study with the commercial solver Gurobi on a workstation.

\subsection{Experimental Setup}
The deterministic case includes 3 CDCs, 15 heterogeneous GPU nodes, 103 LLM training tasks, and a 24-h horizon with one-hour time slots. Each CDC has 4 thermal zones, and the workload contains 72 NTS tasks, 30 TS tasks, and 1 already-running task. Node types are listed in Table~\ref{table:node_specs}. Idle power is set to 20\% of rated node power. Power is represented in kW; under the one-hour case assumption, slot energy is in kWh and cumulative energy is reported in MWh. The formulation represents independently executable, single-node training jobs; it does not model synchronized distributed training, checkpoint failures, or communication collectives during execution. The supplied package does not document measurement provenance for every task and facility parameter, so the study is treated as a parameterized case rather than a production trace.

We conduct three experiments using identical tasks, nodes, CDC assignments, rewards, hold/abort penalties, transfer costs, and energy parameters. The \emph{Proposed} model retains cross-CDC transfer and task-reordering flexibility. The \emph{No Transfer (No TF)} variant prohibits cross-CDC transfer while preserving reordering. The \emph{No Transfer with FIFO-prefix order (No TF+FIFO)} variant additionally prevents a later task from being selected when an earlier task in the same fixed order is skipped. This restriction is stronger than a general operational FIFO policy. Proposed versus No TF isolates cross-CDC transfer within this model, whereas No TF versus No TF+FIFO isolates relaxation of the implemented FIFO-prefix restriction. Monetary quantities are reported in NMU, energy prices in NMU/kWh, and transfer quantities in NDU or NDU/slot as applicable. These same-instance comparisons are descriptive, not statistical.

The Proposed case was solved with Gurobi 11.0.3 on an AMD Ryzen 9 3900X using 24 threads. Before presolve, the model contained 259,433 constraints, 231,097 variables, and 1,122,754 nonzero coefficients. After 14,264.21 s, Gurobi returned an incumbent of 26,025.1126 NMU, a best bound of 26,027.0789 NMU, and a 0.0076\% relative gap under the configured $10^{-4}$ tolerance. Equivalent terminal logs were not available in the supplied package for the two restricted variants; their results are therefore reported without a solver-quality equivalence claim.

\begin{table}[t]
\caption{Core fixed-instance comparison. Profit is in normalized monetary units (NMU), and TF denotes transferred tasks.}
\label{table:core_comparison}
\centering
\scriptsize
\begin{tabular}{|l|r|c|c|c|c|c|}
\hline
\textbf{Model} & \textbf{Profit} & \textbf{Done} & \textbf{Rate} & \textbf{TF} & \textbf{Abort} & \textbf{Hold} \\
\hline
Proposed & 26,025.11 & 103 & 100.0\% & 39 & 0 & 0 \\
No TF & 16,483.81 & 76 & 73.8\% & 0 & 6 & 21 \\
No TF+FIFO & $-$8,629.92 & 14 & 13.6\% & 0 & 30 & 59 \\
\hline
\end{tabular}
\end{table}

\begin{table}[t]
\caption{Scheduling behavior and service capability. J/S denotes concurrent jobs/active servers.}
\label{table:scheduling_comparison}
\centering
\scriptsize
\begin{tabular}{|l|r|c|c|c|c|}
\hline
\textbf{Model} & \textbf{Transfer} & \textbf{Mean} & \textbf{Peak} & \textbf{Mean} & \textbf{CDC Done} \\
 & \textbf{Data} & \textbf{Jobs} & \textbf{J/S} & \textbf{Finish} & \textbf{0/1/2} \\
\hline
Proposed & 320.62 & 10.54 & 11/11 & 13.32 & 45/23/35 \\
No TF & 0.00 & 6.92 & 8/10 & 12.51 & 51/13/12 \\
No TF+FIFO & 0.00 & 1.00 & 1/1 & 12.50 & 11/2/1 \\
\hline
\end{tabular}
\end{table}

\begin{table}[t]
\caption{Energy comparison. Grid, charge, and discharge are in MWh; peak is in kW. U/O denotes node-utilization/server-on percentages.}
\label{table:energy_comparison}
\centering
\scriptsize
\begin{tabular}{|l|r|r|r|r|c|}
\hline
\textbf{Model} & \textbf{Grid} & \textbf{Peak} & \textbf{Charge} & \textbf{Disch.} & \textbf{U/O} \\
 & \textbf{(MWh)} & \textbf{(kW)} & \textbf{(MWh)} & \textbf{(MWh)} & \textbf{(\%)} \\
\hline
Proposed & 2.145 & 137.23 & 0.579 & 1.231 & 70.3/70.3 \\
No TF & 5.882 & 357.46 & 0.814 & 1.479 & 46.1/52.2 \\
No TF+FIFO & 0.000 & 0.00 & 0.367 & 1.006 & 6.7/6.7 \\
\hline
\end{tabular}
\end{table}

\subsection{Comparative Results}

\subsubsection{Flexibility, Profit, and Completion}
In this fixed instance, Table \ref{table:core_comparison} shows monotonic improvement as the two modeled flexibilities are restored. Enabling cross-CDC transfer increases profit from 16,483.81 to 26,025.11 NMU, a gain of 9,541.30 NMU or 57.9\% relative to No TF, while completing 27 additional tasks. Relaxing the FIFO-prefix restriction has a larger descriptive difference: No TF+FIFO completes 14 tasks and yields negative profit because 30 TS tasks are aborted and 59 NTS tasks are held, whereas No TF completes 76 tasks. The proposed model completes all 103 tasks without aborting or holding any task. These comparisons quantify the two flexibility effects for the stated formulation and instance; they do not establish their distribution across other workloads.

\subsubsection{Scheduling Behavior and Service Capability}
Table \ref{table:scheduling_comparison} reports the associated scheduling behavior. The proposed model sustains 10.54 concurrent tasks on average, compared with 6.92 for No TF and 1.00 for No TF+FIFO. Its peak task/computing-node concurrency reaches 11/11, and completed tasks are distributed as 45/23/35 across the three CDCs. The optimized transfer of 320.62 normalized data units accompanies the increase in completed-task throughput rather than merely changing execution locations. Although the mean completion slots are similar across the variants, the proposed model completes substantially more tasks within the same 24-slot horizon.

\subsubsection{Grid Energy, Server Activity, and Battery Operation}
Table \ref{table:energy_comparison} reports grid purchases, battery operation, and node activity while excluding electricity-purchase cost. Despite completing 27 more tasks, the proposed solution purchases 2.145 MWh from the grid, approximately 3.736 MWh (63.5\%) less than No TF based on the unrounded values. This is a grid-purchase comparison, not a general energy-efficiency result: the variants complete different numbers of tasks and use storage differently.

\begin{figure*}[!t]
    \centering
    \includegraphics[width=\textwidth]{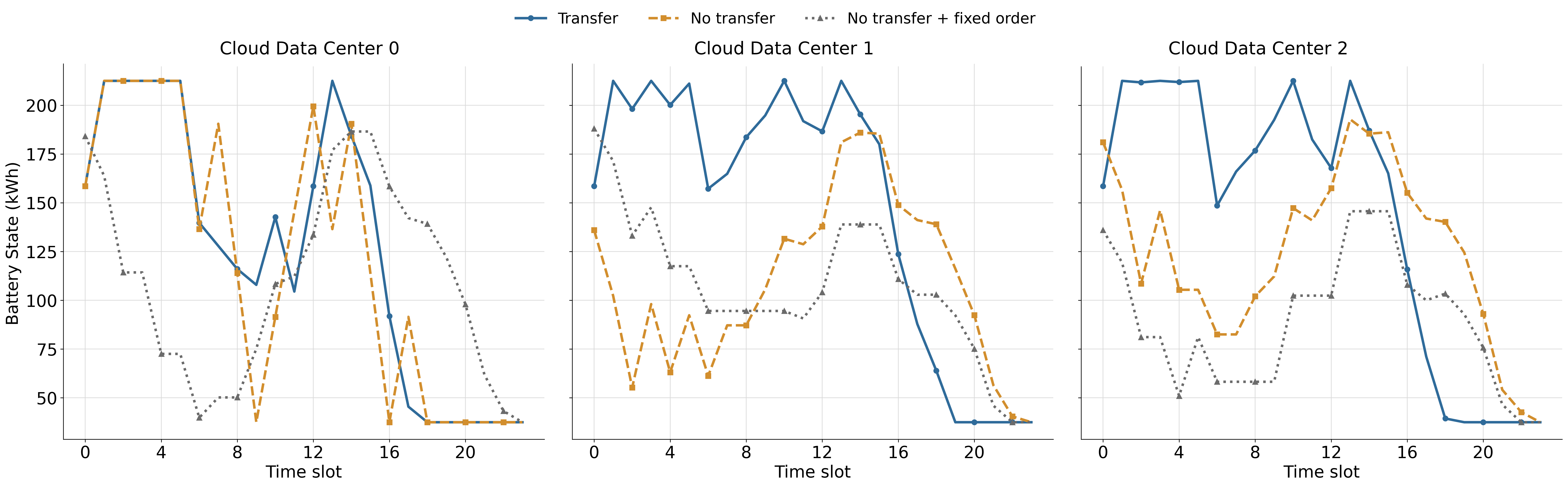}
    \caption{Battery-state trajectories in kWh across the three CDCs for the proposed and baseline scheduling variants.}
    \label{fig:battery_state}
\end{figure*}

Figure \ref{fig:battery_state} compares stored-energy trajectories across the three CDCs. The proposed trajectory declines from 158.5 kWh at the first recorded slot to 37.5 kWh at the last recorded slot in each CDC. Because the supplied package does not document a cyclic terminal-storage requirement, part of the 24-h grid-purchase outcome depends on using initially available stored energy. No TF relies more heavily on CDC 0, whereas No TF+FIFO combines low grid purchase with low activity. Its zero grid purchase does not imply greater energy efficiency because it completes only 14 tasks and activates computing nodes in 6.7\% of node-time slots.

\subsubsection{Detailed Scheduling and Node Selection}
Figure \ref{fig1:task_state} presents the proposed task schedule, where \textit{jp}, \textit{tjp}, and \textit{wj} denote NTS, TS, and already-running tasks. The implemented inventory contains the five node types in Table \ref{table:node_specs}. DGXH100, DGXA200, and DGXA100 nodes carry most of the displayed schedule, whereas the DGX1 nodes remain off. Because many high-reward tasks originate in CDC 0, the optimized solution assigns selected tasks to remote nodes when the resulting transfer, execution, and site-operation decisions improve the modeled objective. The schedule therefore reflects the joint formulation; it does not by itself establish a general performance-per-watt ranking among node types.

\begin{figure}[!t]
    \centering
    \includegraphics[width=0.94\linewidth]{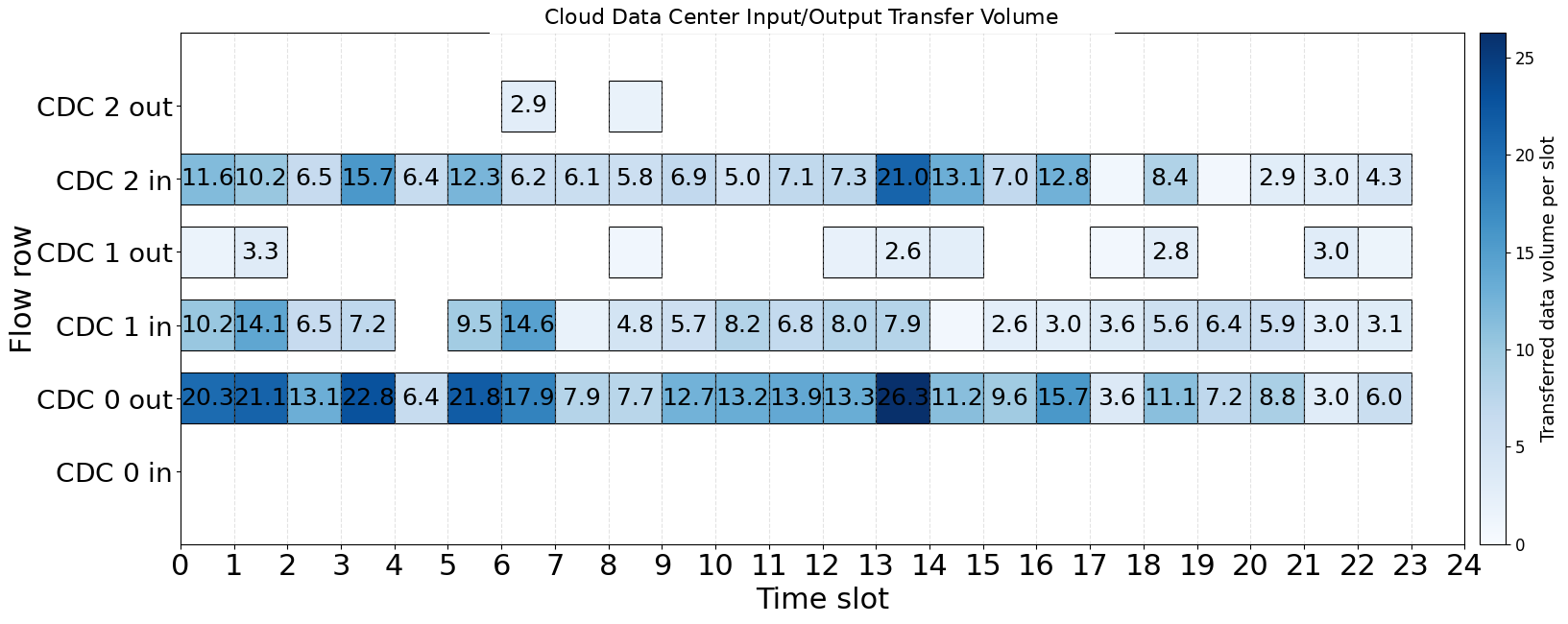}
    \caption{Cross-CDC inbound and outbound transfer decisions.}
    \label{fig3:io_transfer}
\end{figure}

\subsubsection{Cross-CDC Transfer Scheduling}
Figure \ref{fig3:io_transfer} visualizes the traffic exchanged among the three CDCs and its temporal evolution. The optimized schedule has a total outbound transfer volume of 320.62 normalized data units and remotely executes 39 tasks. The transfer pattern is asymmetric: 14 tasks move from CDC 0 to CDC 1, 19 from CDC 0 to CDC 2, 5 from CDC 1 to CDC 2, and 1 from CDC 2 to CDC 1. The solution allocates communication capacity selectively when remote placement improves the modeled objective. It also schedules data movement near the corresponding execution windows while satisfying the modeled inbound and outbound bandwidth limits. These observations describe how transfer flexibility is used in this instance without implying that the same pattern holds for other workloads.

\subsubsection{Thermal and Air-Conditioning Management}
The optimized zone and air-conditioning trajectories remain within the reported modeled limits throughout the 24-h horizon. Zone temperatures lie between 24$^\circ$C and 27$^\circ$C, and the final average zone temperature of each CDC is 24$^\circ$C. AC supply temperatures lie between 24$^\circ$C and 25.81$^\circ$C. These extrema establish feasibility with respect to the modeled temperature constraints; they do not establish robustness to thermal-parameter error or production safety margins. The result nevertheless illustrates why node placement and facility operation are evaluated together: a compute assignment changes the heat input assigned to its zone and the corresponding modeled cooling requirement.

\subsubsection{Summary of Findings}
The comparison identifies complementary effects from the two modeled flexibilities. Relaxing the FIFO-prefix constraint avoids the severe head-of-line blocking observed in that restricted variant, while enabling cross-CDC transfer adds 9,541.30 NMU of profit relative to No TF and expands the feasible placement choices. The proposed case has the highest completion, concurrency, utilization, and profit among the three reported solutions while satisfying the modeled bandwidth and temperature constraints. For this instance, the results support the potential value of evaluating computing, communication, and facility decisions jointly. Broader conclusions require additional workload instances, sensitivity studies, practical scheduling baselines, and consistent solver-quality records for all variants.

%

\section{Conclusion}
\label{par:C}
This paper used a coupled multi-CDC optimization model to separate the fixed-instance effects of cross-CDC task transfer and task-reordering flexibility. Relative to No TF, the proposed model increases profit by 57.9\% and completes 103 rather than 76 tasks. The more restrictive No TF+FIFO variant completes 14 tasks and yields negative profit, exposing the effect of head-of-line blocking under the implemented FIFO-prefix constraint. The proposed schedule transfers 39 tasks and satisfies the modeled bandwidth and zone-temperature limits. These results illustrate the potential value of coordinating computing, communication, and facility operation in the studied instance. Establishing general performance requires repeated workload instances, transfer- and energy-parameter sensitivities, additional scheduling baselines, documented terminal-storage assumptions, and consistent solver-quality statistics for all variants.

\section*{Acknowledgment}
This work was supported by the U.S. National Science Foundation under Award No. 2541996.

\bibliographystyle{IEEEtran}
\bibliography{Solar_ref}

\end{document}